\documentclass[%
 reprint,
 amsmath,amssymb,
 aps,
prl,
]{revtex4-1}

\usepackage{graphicx}
\usepackage{dcolumn}
\usepackage{bm}
\usepackage[hidelinks]{hyperref}
\usepackage{textcomp}
\usepackage[utf8]{inputenc}
\usepackage{color}

\begin{document}

\title{Beyond the Limits of Conventional Stark Deceleration}%

\author{David Reens}
\thanks{dave.reens@colorado.edu}
\altaffiliation{Present Address: Lincoln Laboratory, Massachusetts Institute of Technology, Lexington, Massachusetts 02420, USA}
\thanks{Contributed equally. Email dave.reens@colorado.edu or hao.wu@colorado.edu.}

\author{Hao Wu}
\thanks{hao.wu@colorado.edu}
\altaffiliation{Present Address: Department of Physics and Astronomy, University of California, Los Angeles, California 90095, USA}
\thanks{Contributed equally. Email dave.reens@colorado.edu or hao.wu@colorado.edu.}

\author{Alexander Aeppli}
\author{Anna McAuliffe}

\author{Piotr Wcis\l o}
\altaffiliation{Present Address: Institute of Physics, Faculty of Physics, Astronomy and Informatics, Nicolaus Copernicus University, Grudziadzka 5, PL-87-100 Toru\'n, Poland}

\author{Tim Langen}
\altaffiliation{Present Address: 5. Physikalisches Institut and Center for Integrated Quantum Science and Technology (IQST), Universit\"at Stuttgart, Pfaffenwaldring 57, 70569 Stuttgart, Germany}

\author{Jun Ye}
\affiliation{JILA, National Institute of Standards and Technology and the University of Colorado and\\ Department of Physics, University of Colorado, Boulder, Colorado 80309-0440, USA}

\date{\today}


\begin{abstract}
Stark deceleration enables the production of cold and dense molecular beams with applications in trapping, collisional studies, and precision measurement. 
Improving the efficiency of Stark deceleration, and hence the achievable molecular densities, is central to unlock the full potential of such studies.
One of the chief limitations arises from the transverse focusing properties of Stark decelerators. 
We introduce a new operation strategy that circumvents this limit without any hardware modifications, and experimentally verify our results for hydroxyl radicals. 
Notably, improved focusing results in significant gains in molecule yield with increased operating voltage, formerly limited by transverse-longitudinal coupling. 
At final velocities sufficiently small for trapping, molecule flux improves by a factor of four, and potentially more with increased voltage. 
The improvement is more significant for less readily polarized species, thereby expanding the class of candidate molecules for Stark deceleration.
\end{abstract}

\maketitle

Over the past two decades, Stark deceleration~\cite{VandeMeerakker2008,VanDeMeerakker2012}, where time-varying inhomogeneous electric fields are used to slow polarizable molecules, has enabled groundbreaking collisional~\cite{Sawyer2011,Kirste2012,Gao2018} and spectroscopic~\cite{Veldhoven2004,Hudson2006,Lev2006,Fast2018} studies of a variety of species.
Subsequent trap-loading~\cite{Bethlem2002,Sawyer2007} greatly enhances interrogation time for such studies~\cite{Sawyer2008} and opens the door for further manipulation~\cite{Reens2017}. 
Alongside the history of achievements enabled by Stark deceleration runs a parallel ongoing saga surrounding their efficient operation. 
Many important steps have been made, not only in understanding the flaws of the canonical pulsed decelerator~\cite{VanDeMeerakker2006,Sawyer2008a}, but also in addressing them through the use of overtones~\cite{VanDeMeerakker2005a,Scharfenberg2009}, undertones~\cite{Zhang2016}, or mixed phase angles~\cite{Parazzoli2009,Hou2013}. 
Even with these advances, outstanding inefficiencies of the pulsed decelerator, particularly with regard to transverse phase stability, have motivated alternative geometries such as interspersed quadrupole focusing~\cite{Sawyer2008a} and traveling wave deceleration~\cite{Osterwalder2010,VandenBerg2014,Fabrikant2014}.
Although traveling wave deceleration takes a strong step toward truly efficient operation, it comes with significant engineering challenges. 
These may be partially addressed by the combined use of pulsed and traveling wave devices~\cite{Quintero-Perez2013}, or using traveling wave geometry with pulsed electronics~\cite{Hou2016,Shyur2017}. 
In Zeeman deceleration, the magnetic analog of Stark deceleration, early demonstrations~\cite{Vanhaecke2007,Narevicius2008} were later improved through the use of anti-Helmholtz configurations with better transverse focusing properties~\cite{LavertOfir2011,Dulitz2014}.
Lacking a comparable breakthrough for Stark devices, others have resorted to brand new geometries~\cite{Wang2016}, or combined the Stark and Zeeman approaches in a single device~\cite{Cremers2017,Plomp2019}.

In contrast, we present a new strategy for Stark decelerators that works with conventional geometry and electronics. 
Our strategy fully resolves transverse challenges and improves yields at all final speeds.
It is readily applicable to existing decelerators and thus promises improvements in fields ranging from collisional studies and molecular trapping to precision measurements~\cite{Aggarwal2018}.

\begin{figure*}[t!]
\includegraphics[width=\linewidth]{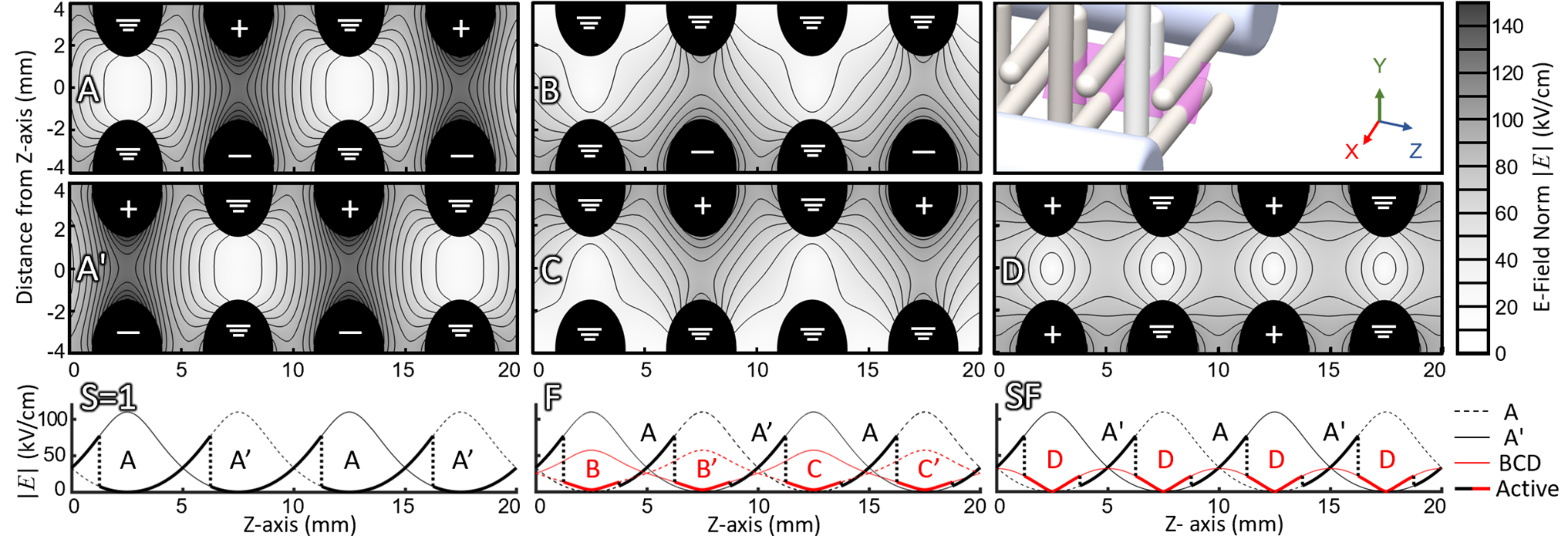}%
\vspace{-2mm}
\caption{
Beyond the limits: conventional Stark deceleration (left column) and transversely focusing variants (middle, right columns).
Conventional Stark deceleration utilizes a single distribution of electric field (A, top) and its translation plus $90^\circ$ rotation (A$'$, middle) in an alternating fashion termed ``S\,=\,1'' operation (bottom, aligned)~\cite{VanDeMeerakker2005a}.
The bottom panel describes S\,=\,1 by plotting the potential experienced by an ideal ``synchronous molecule'' as it propagates exactly down the center axis of the decelerator (bold).
Abrupt changes (bold dashed) in the potential the synchronous molecule experiences are achieved by rapidly switching between A and A$'$ (labels indicate where each is active) via fast high voltage switches.
Distribution A scarcely focuses, and is not active where it most strongly focuses ($z=$7.5~mm, note increasing field strength off-axis).
Distributions are shown in the diagonally slicing plane visible in the 3D render (top right, pink).
A new focusing mode (F, middle column) circumvents this focusing limitation through the incorporation of new distributions B and C (and their primes, unshown, which relate as do A and A$'$).
These distributions do not focus on their own but only when averaged together.
A and A$'$ are still used close to 5~mm and 10~mm as in S\,=\,1, with the result that the energy removed per pin pair (total length of bold-dashed lines) is equivalent to S\,=\,1.
Mode SF (strong focusing, right column) simply replaces B, C, and their primes with D, which is more strongly focusing but challenging to implement experimentally. 
\vspace{-4mm}
}
\label{fig:chargecartoon}
\end{figure*}

To understand this strategy, we revisit the operating principles of a Stark decelerator.
The conventional pulsed Stark decelerator consists of an electrode array with alternating pairs of pins orthogonal to a beamline that passes between them, see Fig.~\ref{fig:chargecartoon}, top right for a 3D render.
In the conventional S\,=\,1 strategy~\cite{VanDeMeerakker2012}, as low-field seeking molecules~\footnote{For strong field seekers, see~\cite{Tarbutt2004}} approach a charged pin pair, they are polarized by the strong electric field and exchange kinetic energy for internal potential energy, effectively climbing a potential hill.
The strong field is then abruptly removed by high voltage switches before the molecules have a chance to regain kinetic energy (Fig.~\ref{fig:chargecartoon}, bottom row).
It is customary to discuss the behavior of an idealized ``synchronous molecule'' that travels along the decelerator axis with zero transverse velocity.
The switching is timed so that the synchronous molecule loses some fixed energy per switch.
It is essential that the synchronous molecule travel only partway up each hill, so that molecules that are ahead of the synchronous molecule get more energy removed, and vice versa.
This generates a longitudinal restoring force for the ensemble, centered on the synchronous molecule.
Transversely, restoring force is not inherited from switching events but arises from the focusing properties of the electric field distributions that the electrode array generates (Fig.~\ref{fig:chargecartoon}, top and middle rows).
When molecules reside in a region where the electric field is stronger off axis than on, they experience transverse focusing.
Although transverse focusing varies rapidly with longitudinal coordinate in the decelerator, these variations are too fast for molecules to follow.
We may therefore make a high speed approximation, and time-average transverse and longitudinal forces to obtain a ``traveling trap'' for the molecules~\cite{Bethlem2000}, which translates along the device and decelerates according to a ramp of the switching frequency.
This is valid provided that $v_z/D \gg f$; $v_z$ the longitudinal velocity of the molecules, $D$ the distance between pin pairs, and $f$ the oscillation frequency in the traveling trap.

Conventionally, pins are always charged in bipolar pairs, in which case transverse focusing occurs between the charged pin pair, but not significantly elsewhere (Fig.~\ref{fig:chargecartoon}(A)).
Molecules do not regularly access the focusing region, since pins are grounded before the synchronous molecule reaches them as discussed above.
As the molecules pass between grounded pins, the transverse field is actually slightly defocusing~\footnote{This is not highly apparent in Fig.~\ref{fig:chargecartoon}A and A', where a $45^\circ$ slicing plane is chosen for visual clarity. The defocusing is strongest in the plane including the decelerator axis that is also normal to the cylindrical axis of the grounded pins.}.  Their transverse confinement also varies with how strongly the molecules are decelerated, and with their distance from the synchronous molecule along the decelerator axis. Such a dependence of transverse confinement on longitudinal position is known as transverse-longitudinal coupling, and it gives rise to the situation that molecules which are coldest longitudinally are less well confined transversely~\cite{VanDeMeerakker2006}.
The use of deceleration overtones such as S\,=\,3~\cite{VanDeMeerakker2005a} alleviates	coupling by allowing molecules to fully transit between charged pin pairs regardless of their relative position with the synchronous molecule. This mode of operation leverages the full focusing properties of the conventional field distribution, but at the expense of only using 1/3 of the pin pairs for removing energy.

Our strategy is to introduce new field distributions with strong transverse restoring forces when the synchronous molecule is between grounded pin pairs, but retain the use of the conventional distribution otherwise. 
Field distributions that focus between grounded pins can be created by charging the neighboring pins to voltages that do not sum to zero.
Field lines then extend toward the grounded pin pair, creating a focusing 2D quadrupole structure.
Possibilities include charging only a single pin as in Fig.~\ref{fig:chargecartoon}(B,C), or charging both to the same voltage, Fig.~\ref{fig:chargecartoon}(D).
We name the operating modes employing these distributions focusing (F) and strong focusing (SF).
We restrict attention to distributions that make use of the same triplet of voltages ($0$, $\pm$12.5\,kV for our device) that are applied conventionally, but rearranged.

\begin{figure*}[t]
\includegraphics[width=\linewidth]{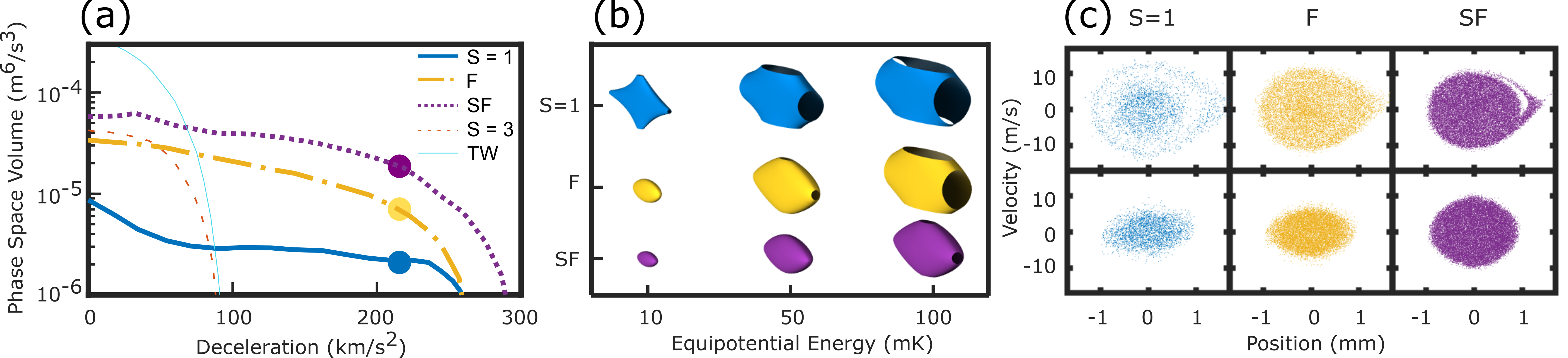}%
\vspace{-2mm}
\caption{
Simulation results of different deceleration modes. 
(a) Simulated phase space volume captured by different modes of operation, for varying decelerations and elapsed time fixed at $3$~ms. 
A $10$~kV peak to peak traveling wave (TW) deceleration and $\pm 12.5$~kV S\,=\,3 are also plotted for comparison. 
Three solid dots correspond to the deceleration used in panels (b) and (c), about $200 \text{ km/s}^2$. 
(b) Equipotentials of the traveling trap generated for three modes. 
Lack of closure of an equipotential indicates the possibility of molecule escape. 
(c) Phase Space Fillings, both longitudinal (top) and transverse (bottom), for the labeled operation modes after $3~\text{ms}$ of travel. 
The surviving number of molecules is $3$, $11$ and $24$ thousand respectively. 
Note dramatic improvements in homogeneity and flux, without significant broadening to larger velocity classes.
\vspace{-4mm}}
\label{fig:efftrap}
\end{figure*}

In order to best compare these modes in a device-independent way, we perform simulations with fixed travel time ($3$~ms) and varying deceleration rate, see Fig.~\ref{fig:efftrap}(a).
S\,=\,1 delivers the smallest phase space volumes, although provides at least some flux even at high deceleration.
Remarkably, F mode offers comparable phase space volume to S\,=\,3, but with triple the deceleration.
SF mode makes more dramatic improvements, extending significant gains to even higher decelerations than possible with any other studied modes.
For the traveling wave (TW) decelerator comparison in Fig.~\ref{fig:efftrap}(a), $10$~kV sine waves are assumed, to our knowledge the largest used to decelerate molecules to rest~\cite{Quintero-Perez2013}. TW offers a good phase-space volume but is limited to a smaller maximum deceleration, similar to S\,=\,3.  
All modes besides TW use the rather small $2$x$2\text{ mm}^2$ open area of our device, while TW devices use rings of $4$~mm inner diameter.
With a $3\times3\text{ mm}^2$~\cite{Scharfenberg2009} or a $4\times4\text{ mm}^2$~\cite{VandeMeerakker2005} device, phase space volume would increase.
Unlike TW however, the performance of F, SF, and S\,=\,1 all degrade significantly when $v_z < 50$~m/s and the high speed approximation breaks down.

In understanding the mechanism for this improved performance, it is helpful to visually inspect the traveling trap generated by each mode, see Fig.~\ref{fig:efftrap}(b).
Here we plot equipotential surfaces for these traps at three different energies and for $200 \text{ km/s}^2$ deceleration.
The openings in these surfaces occur when the surface reaches the $2 \times 2\text{ mm}^2$ transverse limits of our decelerator geometry. Molecules reaching this boundary are lost.
Molecules may also be lost longitudinally, often remaining transversely focused but no longer decelerating with the synchronous molecule.
For S\,=\,1 mode, the $10$~mK equipotential is transversely broad and even contains four small openings.
This corresponds to the  transverse-longitudinal coupling problem discussed above.
The improvements in operation efficiency for F and SF modes correspond to improved tightness and closure as evident in all equipotentials shown.

In Fig.~\ref{fig:efftrap}(c), the longitudinal and transverse phase space fillings are compared for all modes, with $200\text{ km/s}^2$ deceleration and $3$~ms travel time as before. 
All modes are initialized with the same homogeneous phase space density (PSD).
This is valid when the initial beam source generates a much broader distribution than the volume accepted by the traveling trap.
In the longitudinal direction, most supersonic expansions satisfy this, with the exception of those performed with a Helium buffer gas, which can reach temperatures as low as $40\text{ mK}$ expanding from room temperature~\cite{Even2014}.
As can be seen, the distribution is nearly homogeneous after deceleration for all modes except S\,=\,1.
Increases in point density from S\,=\,1 to F, and to SF arise from increases in the phase space volume captured by those operating modes, which is then projected onto the planes shown.
Phase space density is not enhanced, nor could it be by the reversible, non-dissipative Stark deceleration technique.
However, preparing an optimally shaped distribution minimizes subsequent losses in phase space density arising from potentially poor mode-matching.
For example, a trap with an acceptance comparable to the outer dimensions of the S\,=\,1 mode will be under-filled by the S\,=\,1 mode due to the prominent missing ring, while the F mode will not do this, effectively quadrupling the phase space density loaded in such a trap.
Most realistic traps possess comparable transverse and longitudinal phase space acceptances due to ergodicity and cross-dimensional couplings. SF mode is appealing in this respect with nearly identical transverse and longitudinal acceptance. 

\begin{figure}[t]
\includegraphics[width=\linewidth]{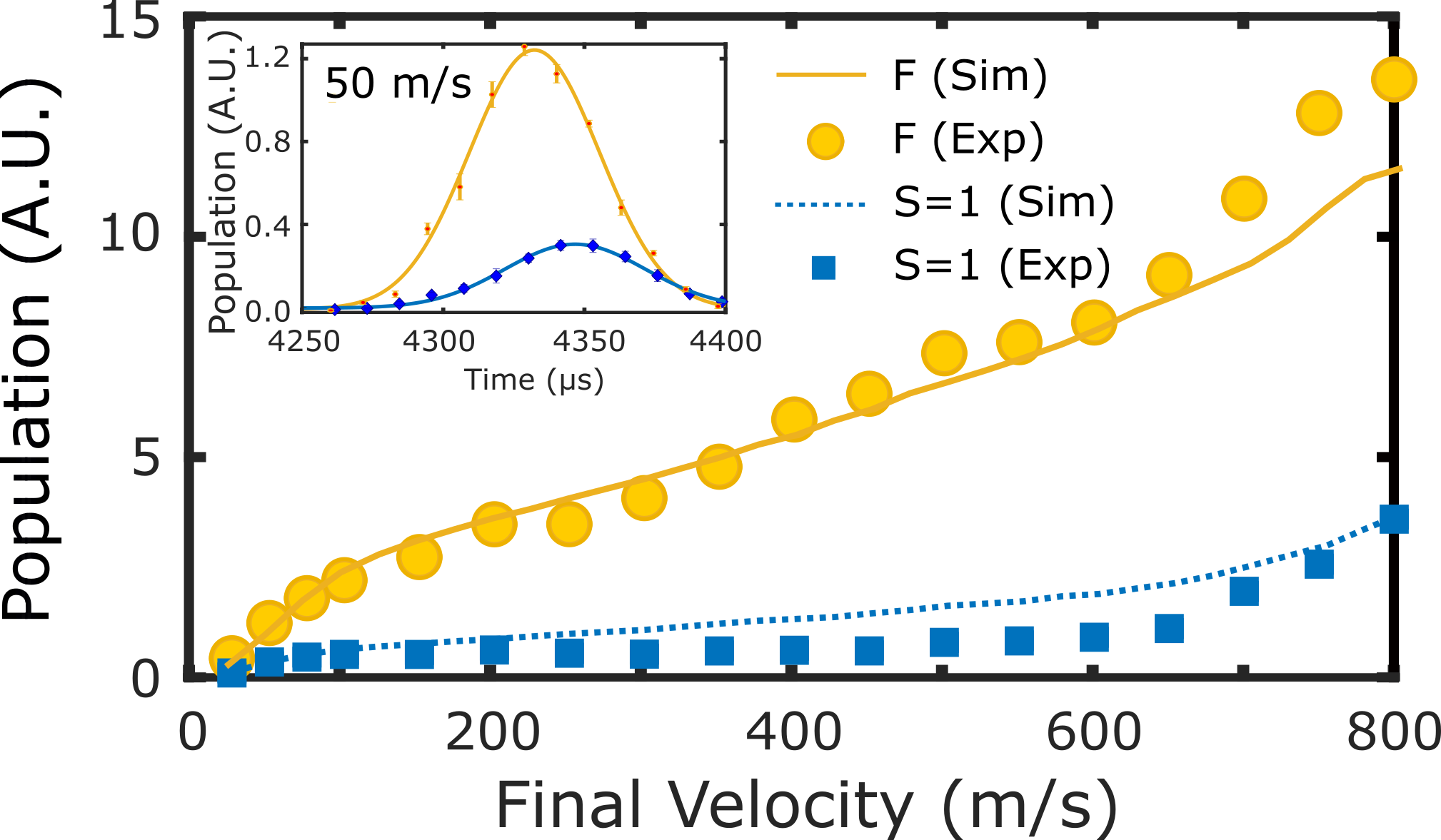}%
\vspace{-5pt}
\caption{\label{fig:alldata}
The molecular signal and enhancement between F mode and conventional S=1 deceleration over a range of final speeds. 
Data are collected with a beam of hydroxyl radicals expanded in neon at an initial speed of $825 \text{ m/s}$ and deceleration up to $200 \text{ km/s}^2$. 
The inset shows the time of flight signal from the valve pulse for F (orange) and S\,=\,1(blue) modes measured at the end of the decelerator when slowing to $50 \text{ m/s}$, demonstrating a factor of $4$ improvement at trappable final speeds. Here the decelerator voltage is 12.5 kV. 
\vspace{-4mm}}
\end{figure}

We experimentally measure the performance of F and S\,=\,1 for hydroxyl radicals, see Fig.~\ref{fig:alldata}.
If the distributions shown in Fig.~\ref{fig:chargecartoon}(A-C) are properly arranged, F mode may be implemented from S\,=\,1 simply by turning off one pin in a pair earlier than the other, and cycling which pin is chosen. 
SF mode requires one electrode to be brought to three different voltages at different parts of the sequence, which remains beyond the capability of fast high voltage switches despite our best efforts.
Data are collected with a beam seeded in neon and an initial speed of $825$~m/s, and run times ranging from $2-4$~ms as the molecules are slowed close to rest.
Pin spacing and most other device parameters are as previously reported~\cite{Bochinski2004,Sawyer2007}, but with increased length.
The signal in S\,=\,1 mode declines rapidly with reduced final speed, but then plateaus, indicative of the improved focusing with stronger deceleration for this mode.
In F mode signal decreases much more gradually, quadrupling S\,=\,1 at the lowest and highest final speeds but improving by more than an order of magnitude in the central $400-500$~m/s range.
For low final velocities below $50\text{ m/s}$ that are used for trap loading, performance enhancement with F becomes particularly attractive as the molecules benefit from a final focusing pulse on the very last pin pair.

\begin{figure}[t]
\includegraphics[width=\linewidth]{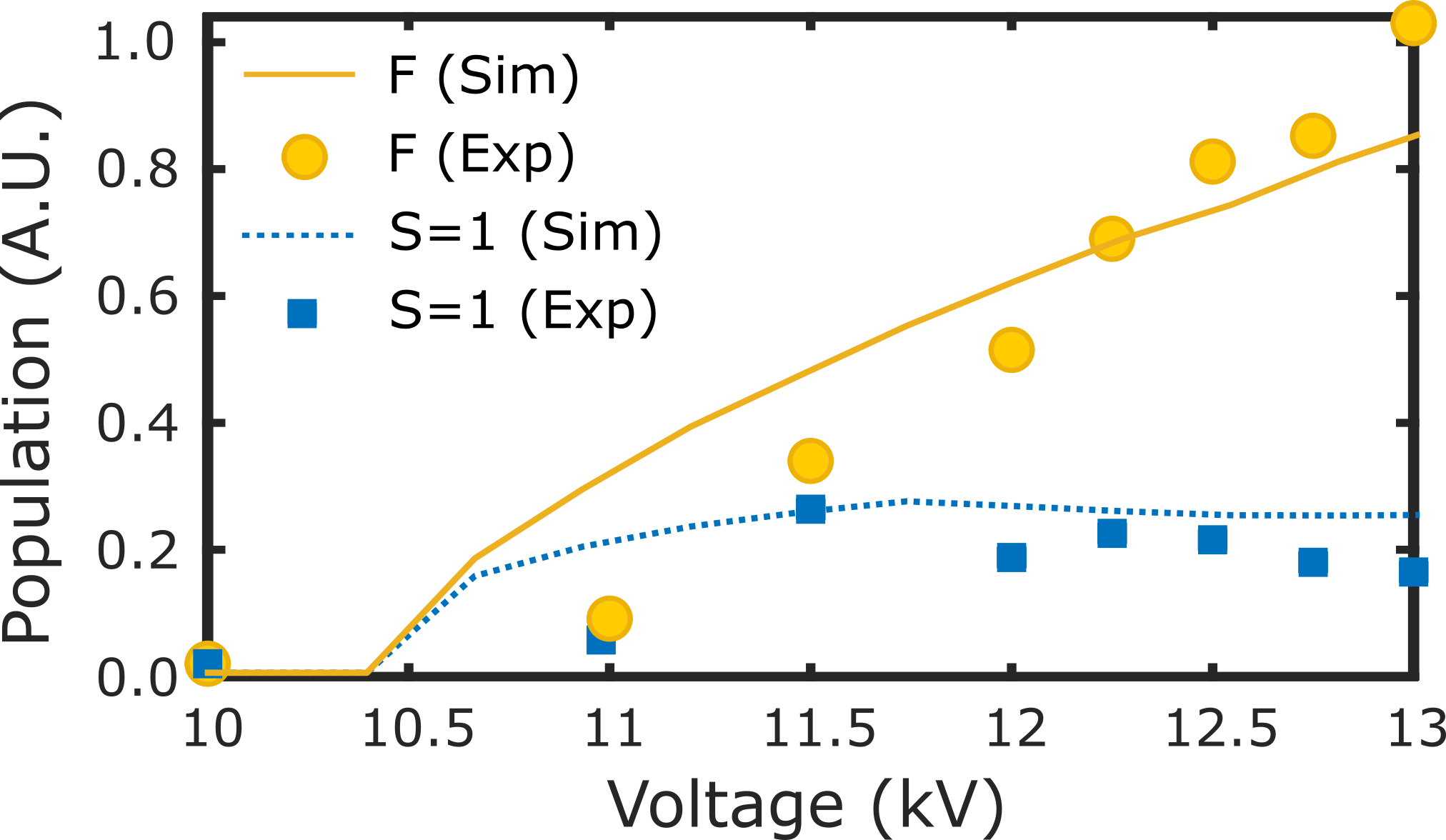}
\vspace{-15pt}
\caption{\label{fig:voltage}
Comparisons of decelerated populations between F mode and S\,=\,1 mode at different applied voltages with a final velocity $50 \text{ m/s}$. 
The points represent experimental results, while the lines are calculated via Monte Carlo simulation. 
Instead of showing saturation behavior as S\,=\,1, the decelerated population using F mode increases with higher applied voltage.
\vspace{-15pt}}
\end{figure}

A particularly direct demonstration of the improved transverse focusing of F mode results from varying decelerator voltage, as done for S\,=\,1 in~\cite{Sawyer2008a}, Fig.~4. 
The hydroxyl radical has a linear Stark shift in our field strengths, so adjusting the voltage linearly scales the potential it experiences.
For operation modes with transverse focusing that is decoupled from longitudinal, voltage increase should only improve performance, deepening the traveling trap.
Figure~\ref{fig:voltage} shows the final population of molecules slowed using S\,=\,1 and F modes to $50 \text{ m/s}$ under different decelerator voltages.
At low enough voltages, the field between the pins is not sufficient to remove enough energy per stage, and molecules cease to be decelerated.
As voltage increases, molecules slowed in S\,=\,1 mode do not need to approach the pins as closely, reducing the sampling of the inter-pin focusing field and worsening performance.
Since the F mode separates transverse focusing from slowing, molecules experience greater transverse focusing at higher field strengths, giving rise to the observed approximately linear improvement above $11$~kV. 
While we are currently limited to $13 \text{ kV}$ by the safety margins of our device, efficiency gains and greater phase space acceptances should persist at even higher voltages, until the initially populated phase space distribution becomes the limitation.
At this point, skimmer cooling~\cite{Segev2017,Wu2018} offers further benefits.

We introduce a new deceleration strategy, with two accompanying modes of operation for the conventional pulsed decelerator. 
Significant improvements in overall performance are demonstrated.
In contrast to deceleration in S\,=\,1 mode, transverse focusing is directly applied by dedicated field distributions with much less dependence on the longitudinal coordinate, enabling further performance gains with increased voltage.
The removal of this dependence also resolves openings in the  traveling trap which previously resulted in significant losses.
This opens up possibilities for applying Stark deceleration to faster beams or to molecules with less favorable Stark shift to mass ratios, since decelerator length may be extended without suffering from increased loss due to greater time spent in traveling traps with openings.
In addition to the two new operation modes identified here, a whole class of deceleration modes incorporating new field distributions is ready for exploration.

We acknowledge funding support for this work from NIST, ARO-MURI, and NSF Grant No. PHY-1734006. 
P. Wcis\l o acknowledges support from the Polish Ministry of Science and Higher Education. 
T.Langen acknowledges support from the Alexander von Humboldt Foundation through a Feodor Lynen Fellowship.

\bibliographystyle{apsrev4-1_no_Arxiv}
\bibliography{allrefs}

\end{document}